\documentclass[sigconf]{acmart}
\AtBeginDocument{%
  }

\begin{document}

\title{Bridging the Gap: Adapting Evidence to Decision Frameworks to support the link between Software Engineering academia and industry}

\author{Patrícia G. F. Matsubara}

\orcid{0000-0001-9230-3620}
\affiliation{%
  \institution{Federal University of Mato Grosso do Sul (UFMS)}
  \city{Campo Grande}
  \country{Brazil}
}
\email{patricia.gomes@ufms.br}

\author{Tayana Uchôa Conte}

\affiliation{%
  \institution{Federal University of Amazonas (UFAM)}
  \city{Manaus}
  \country{Brazil}}
  \email{tayana@icomp.ufam.edu.br}
\orcid{0000-0001-6436-3773}

\renewcommand{\shortauthors}{Matsubara and Conte}

\begin{abstract}
Over twenty years ago, the Software Engineering (SE) research community have been involved with Evidence-Based Software Engineering (EBSE). EBSE aims to inform industrial practice with the best evidence from rigorous research, preferably from systematic literature reviews (SLRs). Since then, SE researchers have conducted many SLRs, perfected their SLR procedures, proposed alternative ways of presenting their results (such as Evidence Briefings), and profusely discussed how to conduct research that impacts practice. Nevertheless, there is still a feeling that SLRs' results are not reaching practitioners. Something is missing. In this vision paper, we introduce Evidence to Decision (EtD) frameworks from the health sciences, which propose gathering experts in panels to assess the existing best evidence about the impact of an intervention in all relevant outcomes and make structured recommendations based on them. The insight we can leverage from EtD frameworks is not their structure per se but all the relevant criteria for making recommendations to practitioners from SLRs. Furthermore, we provide a worked example based on an SE SLR. We also discuss the challenges the SE research and practice community may face when adopting EtD frameworks, highlighting the need for more comprehensive criteria in our recommendations to industry practitioners.
\end{abstract}

\ccsdesc[500]{Software and its engineering}
\ccsdesc[500]{Social and professional topics~Management of computing and information systems}

\keywords{Evidence-Based Software Engineering, Evidence to Decision}

\maketitle

\section{Introduction}
\label{sec:intro}

Evidence-Based Software Engineering (EBSE) aims to provide the means by which the best evidence from rigorous research could be used to make decisions in industrial practice \cite{kitchenham_evidence-based_2004}. An implication from this is the requirement of research in Software Engineering (SE) to be relevant for industry practice. Another implication is the need to search for the currently available best evidence. SE researchers achieve this through systematic literature reviews (SLRs), whose goal is ``to search for and identify all relevant material related to a given topic'' \cite[Section 1.3]{kitchenham_evidence-based_2015}. Researchers follow a set of well-defined procedures, which are meant to be ``as objective, analytical, and repeatable as possible'' \cite[Section 1.3]{kitchenham_evidence-based_2015}. 

Nevertheless, finding all the available evidence regarding a topic employing a reliable process does not necessarily mean such evidence is ``credible'', i.e., evidence we can trust \cite{wohlin_challenges_2021} to make decisions and take action in the real world. There might be insufficient evidence to support a strong conclusion, and more research might still be required to raise confidence in one intervention or another. Considering this, Dyba et al. \cite{dyba_strength_2008} introduced the SE research community to the GRADE (Grades of Recommendation, Assessment, Development, and Evaluation) methodology for assessing how ``strong'' is the existing evidence supporting research findings from SLRs.

Still, even if SE researchers find strong and reliable evidence supporting an intervention, it must reach practitioners' ears and bring with it all relevant information for decision-making. In part, this requires presenting actionable results from research and describing the implications of findings for practice. In other words, this requires the making of clear recommendations for practitioners. Curiously, GRADE is not only about the strength assessment of a body of evidence but is also about making and assessing recommendations for practitioners---something that we, as SE researchers, seem to have missed from the seminal paper of Dyba et al. \cite{dyba_strength_2008}. In this paper, we address this gap by briefly summarizing an approach for making such recommendations, thus addressing the connection between academia and industry: Evidence to Decision (EtD) frameworks. We also present a worked example of EtD frameworks and discuss the challenges to its application in the SE domain. In the next section, we present related work.


\section{Background and Related Work}
\label{sec:background}

In 2008, Dyba et al. \cite{dyba_strength_2008} raised concerns about how much confidence one can have in SLR results, introducing the SE community to the need of evaluating the quality of primary studies. Since then, a growing number of SLRs have assessed the quality of primary studies in SE\cite{yang_quality_2021}. Dyba et al. \cite{dyba_strength_2008} also presented the GRADE methodology for assessing the strength of a body of evidence. GRADE suggests that the analysis of the strength of the evidence supporting a research finding is more than just the quality of the primary studies. It should include other relevant dimensions: publication bias, inconsistency of results, imprecision of results, and the indirection coming from differences in population, intervention, and outcome of interest targeted for answering a research question. Issues in these dimensions are reasons to downgrade the confidence of otherwise strong evidence \cite{balshem_grade_2011}. The approach recommends the summarization of research findings and the judgments for each relevant dimension of the strength of evidence in Evidence Profiles (where judgments for all GRADE dimensions are detailed) or Summary of findings (where only the overall judgment for strength is presented) tables \cite{guyatt_grade_2011-6}. However, very few SLR in the SE domain have assessed the strength of the evidence they gathered \cite{matsubara_unfinished_2025}.

Cartaxo et al. \cite{cartaxo_evidence_2016} proposed Evidence Briefings (EBs), brief documents summarizing SLR results, to disseminate SLR' findings to practitioners. Despite being a clear and understandable medium, EBs may not fully answer the practical questions practitioners have \cite{cartaxo_evidence_2016}. Also, EBs may not include all the information required to support a decision from a practitioner in their daily practice, such as the resources needed for adopting a method or tool \cite{ali_is_2016}.

Back to GRADE, the approach also aims to assist the development of guidelines with recommendations for practitioners based on the evidence from SLRs \cite{guyatt_grade_2011-6}. The working group responsible for developing GRADE proposed the idea of Evidence to Decision (EtD) frameworks, which aggregate information (evidence included) to assist practitioners' decision-making regarding interventions. We explain this framework in detail in the next section. 

\section{The elements of an EtD Framework}

\begin{figure*}[htpb]
  \centering
  \includegraphics[width=1\textwidth]{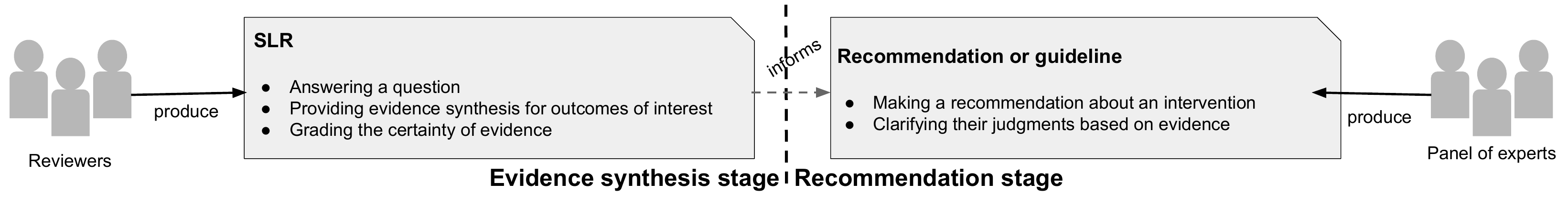}
  \caption{The relationship between SLRs and recommendations/guidelines.}
  \Description[Figure to show how SLRs inform recommendations and guidelines]{The figure shows that researchers conducting a SLR provide findings around outcomes of interest to answer a review question. They also grade the evidence stregnth. All these information informs panel of experts when creating a recommendation or a guideline. The panel of experts should also use the SLR evidence to help clarify their judgments.}
  \label{fig:SLRs}
\end{figure*}

Originally, an EtD is meant to help creating and assessing a recommendation or guideline regarding a clinical question about what to do when treating a patient, considering the evidence about existing interventions. Its creation involves two major stages, as Figure \ref{fig:SLRs} illustrates: an evidence synthesis and a recommendation stage.

During the evidence synthesis stage, reviewers gather the best available evidence to answer the question through typical SLR procedures. This means synthesizing evidence and concluding about the targeted intervention's effects on critical outcomes of interest. Reviewers also critically appraise the confidence (or strength) on such evidence, i.e. on the estimates of the effects of the intervention. 

During the recommendation stage, a panel of experts knowledgeable in SLR procedures, the GRADE system, and the topic of interest are responsible for making the recommendation. They will use the information from the SLR, along with other criteria defined by EtD frameworks to do so. The panel will follow three steps, which correspond to the three sections of the EtD structure \cite{alonso-coello_grade_2016-1}, which we explore in the next sections:

\begin{enumerate}
    \item Formulating the question (Section \ref{subsec:question});
    \item Making an assessment (Section \ref{subsec:assessment}); and
    \item Drawing a conclusion (Section \ref{subsec:recommendation}).
\end{enumerate}

\subsection{Formulating the question}
\label{subsec:question}

The first step in the EtD framework is formulating a question to address. Table \ref{tab:question} presents all the items that a panel of experts must consider when defining the question to structure their judgments and to reach a recommendation. It also shows a worked example of a question for the SE domain on the topic of pair programming in comparison with solo programming. The worked example is inspired by the SLR results of Hannay et al. \cite{hannay_effectiveness_2009}, but not necessarily wholly faithful to the most current and best-existing evidence on the topic. When needed, we also inserted fictitious decisions for the panel to increase the force of the example and provide means for reflection on what we can represent with EtD frameworks.

\begin{table*}[htpb]
    \centering
    \begin{tabular}{|p{2cm}||p{8cm}|p{7cm}|}
        \hline
         \multicolumn{3}{|p{17cm}|}{\textbf{Question:} The question using the PICO (population, intervention, comparison, outcome) format. \textbf{Example:} Should pair programming be adopted in software development projects? } \\ \hline
         \multicolumn{3}{|l|}{\textbf{Question details:}} \\ \hline
         \textbf{(P) Population:}  & The characteristics of the targeted population for the intervention/comparison. & \textbf{Ex.:} Software developers. \\ \hline
         \textbf{(I) Intervention:}  & The intervention that the recommendation focuses on. & \textbf{Ex.:} Pair programming. \\ \hline
         \textbf{(C) Comparison:} & An alternative treatment to the intervention. & \textbf{Ex.:} Solo programming. \\ \hline
         \textbf{(O) Main outcomes:}  & The critical outcomes of interest to analyze when recommending an intervention or its comparison. & \textbf{Ex.:} Duration, effort, quality. \\ \hline
         \textbf{Setting:}  & The characteristics of the environment where the recommendation is to be implemented. & \textbf{Ex.:} A software development organization in general. \\ \hline
         \textbf{Perspective:}  & The perspective taken for the recommendation. It can be of an individual practitioner deciding to adopt the intervention, or of an organization that can also adopt it. & \textbf{Ex.:} The organization/team. \\ \hline
         \textbf{Subgroups:} & Description of subgroups for which the recommendation is likely to vary due to special characteristics they possess, if there is any. & \textbf{Ex.:} Inexperienced software developers. \\ \hline
         \textbf{Conflicts of interest:}  & Any conflict of interest that any panel member might have that can lead to the favouring of the intervention or of the comparison. & \textbf{Ex.:} Mr. James Smith is a member of an organization that provides training for pair programming dynamics. The remaining panel members declared no conflict. \\ \hline
    \end{tabular}
    \caption{Items (from \cite{alonso-coello_grade_2016-1}) for formulating the question, and their descriptions (by the authors).}
    \label{tab:question}
\end{table*}

The question uses the PICO (population, intervention, comparison, outcome) format, and a few additional details \cite{alonso-coello_grade_2016-1}. When defining the population, the panel must clarify any relevant characteristics of the targeted people for the intervention. For example, only people at a high-risk of a catastrophic outcome (e.g. death) might be targeted. Determining how broad is the population or the intervention is a challenging decision when formulating the question, and a practical guideline is to consider variations of magnitudes of effects of the treatment across subgroups of the population or across differences in treatment options \cite{guyatt_grade_2011-5}. The panel should also be specific about the intervention and the comparison, i.e., the alternative treatment. In medicine, this can involve decisions about doses, and a combination of drugs. The outcomes of interest also clarify the variables used in the assessment of the intervention and the comparison. 

Depending on the question, the definition of the setting is crucial. It helps identify constraints that might limit the recommendations, e.g., the wealth of a country can impose limitations on feasible treatments to provide in health care systems. The perspective helps to create a ``practitioner'' focus, clarifying whether the panel has to answer with an individual or group point of view in mind. It will influence the outcomes of interest to analyze and even economic evaluations \cite{brunetti_grade_2013}. For instance, an individual might have low concerns regarding the costs of a treatment if a health system pays for it. However, the cost information will be most important when assuming a health system perspective. The subgroup aids in specifying sub-populations for which recommendations might differ from the overall population, given their specific characteristics and the existing evidence. 

Also, the panel needs to inform whether any of its members have conflicts of interest---for instance, on promoting the intervention or the comparison for any reason. Valid reasons can include participation in organizations that provide tools or training associated with the intervention, for example. Identifying conflicts will aid the panel in dealing with them, either by just declaring the conflict or excluding members from discussions about specific questions \cite{alonso-coello_grade_2016-1}. 

In the worked example, the panel's recommendation is made without specific constraints, applicable to organizations or teams developing software. It notes that the recommendation might differ for inexperienced developers, who may achieve different results than experienced ones. One of the panel members declared a conflict of interest, as he is a member of an organization that provides training in the topic of interest. Therefore, he might have a more favorable view of the effects of the intervention. In summary, the question provides the details to guide the assessment criteria for making a recommendation. The next section presents these criteria and continues with our worked example.

\subsection{Making an assessment}
\label{subsec:assessment}

In the second step, the panel judges a set of criteria relevant to practitioners' decision-making. These criteria vary with the type of decision. In the health domain, types of decisions include (i) clinical recommendations from an individual perspective, (ii) clinical recommendations from a population perspective, (iii) coverage decisions\footnote{For health insurance providers.}, (iv) health system and public health system recommendations/decisions, and (v) diagnostic, screening, or other tests \cite{alonso-coello_grade_2016-1}.

Table \ref{tab:criteria} shows a few of the criteria for two decision types in Alonso-Coello et al. \cite{alonso-coello_grade_2016}: population-based and individual recommendations. Some criteria are type-specific, while others apply to both types. The supplementary material \cite{matsubara_supplementary_2026} shows the judgments of a fictitious panel of experts for some of these relevant criteria from the perspective of organizations or teams pondering whether to adopt pair programming, aligned with the question in Section \ref{subsec:question}. 

\begin{table*}[htpb]
    \centering
    \begin{tabular}{|p{7.5cm}||p{9.5cm}|}
        \hline
        \textbf{Population perspective} & \textbf{Individual perspective} \\ \hline
         Is the problem a priority (from a population perspective)? & Is the problem a priority (from an individual perspective)? \\ \hline
         \multicolumn{2}{|c|}{How substantial are the desirable anticipated effects?} \\ \hline
         \multicolumn{2}{|c|}{What is the overall certainty of the evidence of effects?} \\ \hline
         \multicolumn{2}{|c|}{Does the balance between desirable and undesirable effects favour the intervention or the comparison?} \\ \hline
         What is the certainty of the evidence of resource requirements (costs)? & Does the cost effectiveness of the intervention (the out-of-pocket cost relative to the net desirable effect) favour the intervention or the comparison? \\ \hline
         Is the intervention acceptable to key stakeholders?* & Is the intervention acceptable to patients, their caregivers, and healthcare providers? \\ \hline
    \end{tabular}
    \caption{Examples of criteria for making assessments for population and individual perspectives (from \cite{alonso-coello_grade_2016}).}
    \label{tab:criteria}
\end{table*}


The judgments are guided by the question Section \ref{subsec:question} presents. For instance, for the criterion ``How substantial are the desirable anticipated effects?'' (from Table \ref{tab:criteria}) the panel used the Summary of Findings table with the evidence for all the outcomes of interest from the question: quality, duration, and effort. The judgment for the criterion ``Is the intervention acceptable to key stakeholders?'' also consider the perspective informed in the question: organizations and teams, instead of the perspective of individual developers. The panel could even explore the relevant subgroup included in the question during their judgments by creating a separate Summary of Findings with evidence for this specific subgroup or additional considerations if there is enough evidence that the results for the outcomes of interest were different for these groups.

\subsection{Drawing a conclusion}
\label{subsec:recommendation}

In the third step, the panel draws a conclusion from their judgments for all criteria and deliver an actionable, specific, and clear recommendation for practitioners. \cite{alonso-coello_grade_2016-1}. The panel informs the recommendation's direction: in favor or against the intervention. The panel must also declare the recommendation's strength, which can be either strong---when it is highly likely that a patient will accept the intervention/treatment in any circumstances---or discretionary---when a patient's acceptance of the intervention/treatment will vary depending on the balance between desirable and undesirable effects, and their values and preferences, among other things \cite{andrews_grade_2013}. 


The supplementary material also shows the panel's recommendations for our worked example \cite{matsubara_supplementary_2026}. Such recommendation is discretionary, as different organizations and teams will likely vary in their decision to adopt pair programming for the many reasons explained in the recommendation. For instance, for some organizations, the undesirable effect of higher effort might be unacceptable. The recommendation also includes details that can help practitioners implement the intervention and monitor and evaluate its efficacy. In the worked example, the panel considers it relevant for organizations to have a productivity measurement system in place to help assess the impact of pair programming on productivity. The recommendation also presents further research directions to increase confidence in the evidence or explore gaps likely necessary for better comprehension of the intervention's effects on relevant outcomes of interest. In the worked example, the panel highlights the importance of considering the individual perspective when making decisions by identifying software developers' values and preferences towards working solo or in pairs. The panel also considers the need to perform primary research on the effects of pair programming on other critical and unexplored outcomes of interest, such as learning, especially when onboarding new team members. Moreover, they consider that measuring developers' satisfaction and well-being helps to understand productivity better \cite{forsgren_space_2021}, so it is also a critical outcome to further research.

The recommendation created with an EtD framework will stem from a clear question defined with practitioners in mind. Then, the ETD framework guides the panel in judging the existing evidence, accounting for all relevant issues practitioners will likely consider in making a decision about an intervention. Such evidence is unlikely to come from a single SLR \cite{zhang_using_2019}, so EtD frameworks provide the means for gathering it in a single place. The framework also allows the panel to highlight the strength of their recommendations based on the objective criteria assessed, which is unusual in SLRs and can be a relevant factor for decision-making. Panels are also expected to update EtD frameworks and their recommendations occasionally or when new relevant evidence emerges. This can also benefit practitioners, enabling them to update their knowledge in their fields of expertise more easily.

\section{Envisioned Challenges for SE}

While the adoption of EtD frameworks in SE may present challenges, it also holds the potential to significantly impact the industry. This section delves into the potential hurdles and issues the SE community must resolve to leverage their benefits.

\textbf{Who are the entities entitled to make a recommendation for industry practitioners, if any?} Clinicians can turn to national specialty associations and even worldwide organizations such as WHO (World Health Organization) for guidelines. SE chapters of professional associations such as  ACM SIGSOFT (ACM Special Software Engineering Group) and IEEE TCSE (IEEE Technical Community on Software Engineering) could assume a similar role in Software Engineering by organizing expert panels to develop and issue guidelines. Additionally, invited panels of technical experts could contribute to guidelines published through Special Issues in selected venues. 

\textbf{We need platforms where practitioners can look for answers to their questions.} Cartaxo et al. \cite{cartaxo_evidence_2016} found that most survey participants preferred platforms like StackExchange over static EBs.  This suggests we should design and promote engaging platforms where the information practitioners need is readily available in the form of answers that EtD frameworks provide.

\textbf{What are industry practitioners' different types of decisions and their relevant criteria?} In Section \ref{subsec:assessment}, we present the different types of decisions discussed in EtD frameworks for the health domain. SE researchers must also identify the types of decisions for which SE research can provide evidence, as this connects to the criteria for making a recommendation. SE clearly can make recommendations from an individual perspective---to software developers, for instance. Kitchenham et al. \cite{kitchenham_evidence-based_2004} highlighted the perspectives of project managers interested in results for specific projects and senior managers interested in results for a department or the whole organization. Such perspectives suggest there is likely a need for recommendations for group perspectives, such as teams and organizations, in SE too.  

\textbf{We must aggregate evidence around outcomes of interest.} Identifying different perspectives is paramount to defining relevant criteria for recommendations and clarifying the different outcomes of interests to consider when conducting primary studies. For instance, while individual developers and software teams might focus on maintainability when considering quality as an outcome of interest, stakeholders with business or managerial points of view might focus on user experience or customer satisfaction \cite{green_developer_2024}. Additionally, acceptability and feasibility may vary between group and individual perspectives, depending on the intervention. 

\textbf{We also need to define the equivalent of ``clinical questions''.} Perhaps we can call such questions ``practice questions'' to adapt to the SE context and differentiate them from research questions. This connects with the finding of Cartaxo et al. \cite{cartaxo_evidence_2016} that, though participants might find SLR findings important, not necessarily such findings answer \textit{their} questions. This is critical and current for primary research also. Winters \cite{winters_thoughts_2024} have recently remarked how research in a highly relevant SE venue still addresses questions of no interest to the software industry or provides solutions with no practical context. 


\section{Conclusion}

The thing with EtD frameworks is not their structure per se but all the information they show practitioners need for making meaningful decisions against or in favor of an intervention. Evidence about interventions' effectiveness is insufficient, as the intervention might be too costly to be adopted \cite{ali_is_2016}, or unacceptable to a relevant set of stakeholders. In this paper, we describe the EtD frameworks and present a worked example of their application in the SE domain. We show that EtD frameworks include information lacking in individual SLRs in SE. EtD frameworks can also aid the SE research community in directing research efforts, by clarifying missing evidence to derive guidelines for practitioners or subgroups of the population who can benefit from more specific recommendations. Therefore, EtD frameworks are a promising method to help us close the academia-industry gap.


\begin{acks}
This work was executed with the support of the Federal University of Mato Grosso do Sul (Universidade Federal de Mato Grosso do Sul – UFMS – Brasil), of the brazilian funding agency FUNDECT (Call number 42/2024), and of the Federal University of Amazonas (Universidade Federal do Amazonas - UFAM - Brasil). This study was financed in part by the Coordenação de Aperfeiçoamento de Pessoal de Nível Superior – Brasil (CAPES) – Finance Code 001, and by Amazonas State Research Support Foundation - FAPEAM - through the POSGRAD project 2025/2026. We also would like to thank the financial support granted by CNPq 445029/2024-2, 443934/2023-1, and 314797/2023-8. 
\end{acks}
\bibliographystyle{ACM-Reference-Format}
\bibliography{references}

\end{document}